\documentclass[aps,pre,showpacs,amsmath,amssymb,amsfonts,lengthcheck,longbibliography,superscriptaddress]{revtex4-1}
\usepackage{graphicx}
\usepackage{subfigure}
\usepackage{amsthm}
\usepackage{verbatim}
\usepackage{dcolumn}
\usepackage{bm}
\usepackage{epsf}
\usepackage{color}
\usepackage[colorlinks=true,citecolor=blue,linkcolor=blue,urlcolor=blue]{hyperref}%
\usepackage{xcolor}
\usepackage{dsfont}

\newcommand{\e}[1]{\exp{\left(#1\right)}}

\newcommand{\bla}{bla\\bla\\bla\\bla\\bla}

\newcommand{\be}{\begin{equation}}

\newcommand{\ee}{\end{equation}}

\newcommand{\ba}{\begin{align}}

\newcommand{\ea}{\end{align}}

\newcommand{\bi}{\begin{itemize}}

\newcommand{\ei}{\end{itemize}}

\begin{document}

\title{Fluctuation theorem for irreversible entropy production in electrical conduction}

\author{Marcus V. S. Bonan\c{c}a}
\email{mbonanca@ifi.unicamp.br}
\affiliation{Instituto de F\'isica `Gleb Wataghin', Universidade Estadual de Campinas, 13083-859, Campinas, S\~{a}o Paulo, Brazil}

\author{Sebastian Deffner}
\email{deffner@umbc.edu}
\affiliation{Department of Physics, University of Maryland, Baltimore County, Baltimore, MD 21250, USA}
\affiliation{Instituto de F\'isica `Gleb Wataghin', Universidade Estadual de Campinas, 13083-859, Campinas, S\~{a}o Paulo, Brazil}

\date{\today}

\begin{abstract}
Linear, irreversible thermodynamics predicts that the entropy production rate can become negative. We demonstrate this prediction for metals under AC-driving whose conductivity is well-described by the Drude-Sommerfeld model.  We then show that these negative rates are fully compatible with stochastic thermodynamics, namely, that the entropy production does fulfill a fluctuation theorem.  The analysis is concluded with the observation that the stochastic entropy production as defined by the surprisal or ignorance of the Shannon information does not agree with the phenomenological approach.
\end{abstract}

\maketitle

\section{Introduction \label{sec.intro}}

The only processes that are fully describable by means of traditional thermodynamics are infinitely slow successions of equilibrium states \cite{Callen1985}. While considering such idealized situations is well-suited to formulate universal statements, its practical insight is somewhat limited.  All real processes occur at finite rates,  and therefore entropy is irretrievably lost into the environment.  Historically the first attempt to quantify this entropy ``production'' was developed in \emph{linear, irreversible thermodynamics} \cite{prigogine1961,mazur2011}.  The central assumption of this generalized theory is that fluxes depend  only linearly on the forces driving the system away from equilibrium.  Its results are obtained from a combination of the local equilibrium hypothesis with conservation laws.  The textbook example is heat conduction generated by a temperature gradient, which is described by Fourier's law \cite{zwanzig2001}.

An even more widely-known example of such linear problems is electrical conduction,  that is described by Ohm's law.  Nevertheless, there are still subtleties to be unveiled.  Only recently, we showed in Ref.~\cite{Bonanca2021} that the entropy production in noble metals under AC driving exhibits non-monotonic growth as a function of time. In other words, in electrical conduction the entropy production rate can become negative. Such negative rates occur in situations, in which the external driving is too fast for the system to react, and hence the response lacks behind the overall dynamics. Similar observations have been reported in, e.g., viscoelastic fluids \cite{Williams2007} and single-level quantum dots \cite{thingna2017} under oscillatory driving and in freely expanding ideal gases \cite{chakraborti2021}.  Interestingly, negative entropy production rates in open systems undergoing non-Markovian dynamics are somewhat commonplace \cite{bhattacharya2017,marcantoni2017,xu2018,popovic2018,strasberg2019}.  However,  our analysis is entirely based on the lag of response \cite{zwanzig1961,McLennan1964}, which also gives rise to negative rates in Markovian settings \cite{Bonanca2020a,Bonanca2021}.

Even more remarkably,  negative entropy production rates are not indicative of scenarios operating far from thermal equilibrium, but they can be observed  fully within the regime of linear, irreversible thermodynamics \cite{Bonanca2020a}.  Since this approach fully rests on phenomenological and macroscopic arguments, the considered entropy production is often considered as genuinely thermodynamic.

A more modern approach to nonequilibrium problems is \emph{stochastic thermodynamics} \cite{peliti2021}.  Arguably, the most central notion is the stochastic entropy production, which is defined as a statistical, or rather information theoretic quantity. Whereas in linear, irreversible thermodynamics the entropy production is expressed as a bilinear form of fluxes and forces, in stochastic thermodynamics we have the ``surprisal'', i.e., the logarithm of the system's distribution in state space \cite{seifert2005}.  The main results of stochastic thermodynamics are the fluctuation theorems, which quantify that negative fluctuations of the entropy production are exponentially suppressed \cite{Crooks1999,seifert2005,Esposito2010a}.

The natural question arises, whether the two paradigms are consistent with each other, or rather whether the two ``versions'' of entropy production are equivalent.  In particular, the occurrence of negative rates in irreversible thermodynamics \cite{Williams2007,Bonanca2020a,Bonanca2021} appears incompatible with the strict positivity of the stochastic entropy production rate often discussed in the literature \cite{spohn1978,Prigogine1986,ruelle1996,Maes2000,Belandria2005,Esposito2010b,Esposito2010c,Tome2010,Bauer2016,Brandner2016,
deffner2017,jabraoui2019,Landi2020,Deffner2020EPL}.

In the present letter, we show that the entropy production as defined in linear, irreversible thermodynamics indeed fulfills a fluctuation theorem. For pedagogical reasons and for specificity we focus on electrical conduction in Drude-Sommerfeld metals \cite{Drude1900I,Drude1900II,Sommerfeld1928,bardeen1940} under AC-driving.  Despite its somewhat crude approximations, experiments have shown that the Drude model, combined with Fermi-Dirac statistics, does describe properties of real metals such as gold, copper and silver \cite{Olmon2012,Yang2015} at room temperature and low photon energies. However, the validity of our results is not restricted to electric conduction. Rather, it is easy to see that our analysis remains valid for other situations that can be described with the linear framework.  Finally, we will also see that the irreversible entropy production and the stochastic entropy production are typically different, and that they only become identical in the limit of infinitely slow driving.

\section{Irreversible thermodynamics \label{sec.irrthermo}}

To keep the discussion self-contained, we begin by briefly reviewing the approach, and by establishing notions and notations.  In our analysis, we use irreversible thermodynamics to derive an expression for the entropy production (EP) rate for a  situation in which monochromatic and polarized light is shined on a piece of metal. In this case, the balance equation for the electromagnetic energy density $u_{\mathrm{EM}}(\mathbf{r},t)$ reads, 
\begin{equation}
\frac{\partial u_{\mathrm{EM}}}{\partial t} + \nabla\cdot\mathbf{S} = - \mathbf{J}_{e}\cdot\mathbf{E}\,,
\label{eq.balanceem}
\end{equation} 
where $\mathbf{S}$ is the Poynting vector. Equation~\eqref{eq.balanceem} contains a source term which describes the power density lost to the charge carriers, which is the product of the electrical current density $\mathbf{J}_{e}$ and the electric field $\mathbf{E}$.  

Accordingly, the balance equation for the internal energy density $u(\mathbf{r},t)$ of the charge carriers becomes,
\begin{equation}
\frac{\partial u}{\partial t} + \nabla\cdot\mathbf{J}_{u} = \mathbf{J}_{e}\cdot\mathbf{E}\,,
\label{eq.balanceIE}
\end{equation}
where $\mathbf{J}_{u}$ denotes the flux of internal energy and, due to energy conservation, the source term has the opposite sign. Further assuming a constant number of charge carriers, we also have the balance equation for the entropy density $s(\mathbf{r},t)$,
\begin{equation}
\frac{\partial s}{\partial t} + \nabla\cdot\mathbf{J}_{s} = \dot{\overline{\Sigma}}\,,
\label{eq.balanceEP}
\end{equation}
whose source term is the EP rate we want to obtain. 

Combining Eqs.~\eqref{eq.balanceIE} and \eqref{eq.balanceEP} with the local equilibrium hypothesis (see Eq.~(\ref{eq.balancerel}) and App.~\ref{app.irrtherm}), we obtain,
\begin{equation}
\dot{\overline{\Sigma}} = \mathbf{J}_{u}\cdot\nabla\left( \frac{1}{T}\right)-\mathbf{J}_{n}\cdot\nabla\left( \frac{\mu}{T}\right) + \frac{ \mathbf{J}_{e}\cdot\mathbf{E}}{T}\,,
\end{equation}
whose last term is the contribution from electrical conduction. The quantities $T$ and $\mu$ denote, respectively, the temperature and the chemical potential imposed on the set of charge carriers as scalar fields, and $\mathbf{J}_{n}$ is the particle flux. 

Temperature, chemical potential and pressure gradients are constraint by the Gibbs-Duhem equation \cite{reichl} (see App.~\ref{app.irrtherm}), which implies that under uniform temperature and pressure, the EP rate simply becomes
\begin{equation}
\dot{\overline{\Sigma}} = \frac{ \mathbf{J}_{e}\cdot\mathbf{E}}{T}\,.
\label{eq.epit}
\end{equation}

Additionally, phenomenological linear relations between currents and forces are assumed to hold.  In electrical conduction, such a relation exists in the frequency domain \cite{kubo2012}, 
\begin{equation}
\hat{J}_{e,i}(\omega) = \sum_{j}\sigma_{i j}(\omega) \hat{E}_{j}(\omega)\,,
\label{eq:ohmslaw}
\end{equation}
i.e., between the Fourier transforms of the electrical current, $\hat{J}_{e,i}(\omega)$, and the electric field, $\hat{E}_{j}(\omega)$, along the $i$ and $j$ direction, respectively.  Note that Eq.~(\ref{eq:ohmslaw}) is nothing but  Ohm's law with the conductivity tensor $\sigma_{i j}(\omega)$.

To evaluate the EP rate $\dot{\overline{\Sigma}}$ (\ref{eq.epit}), an expression for the current in time domain is required. The inverse Fourier transform of Eq.~(\ref{eq:ohmslaw}) reads 
\begin{equation}
J_{e,i}(t) = \int_{-\infty}^{t}dt'\,\sum_{j}\Phi_{i j}(t-t')\,E_{j}(t')\,,
\label{eq:ohmtime}
\end{equation}
where $\Phi_{i j}(t)$ denotes the response function \cite{kubo2012}. Equation~(\ref{eq:ohmtime}) describes possible memory effects since current and electric field are not evaluated at the same instant of time \cite{kubo2012,zwanzig1961}. Combining Eq.~(\ref{eq:ohmtime}) and the bilinear form (\ref{eq.epit}) for $\dot{\overline{\Sigma}}$, we obtain 
\begin{equation}
\dot{\overline{\Sigma}} = \frac{1}{T}\,\sum_{i,j} E_{i}(t) \,\int_{-\infty}^{t}dt'\,\Phi_{i j}(t-t')\, E_{j}(t')\,.
\label{eq:eprdelay}
\end{equation} 
This expression contrasts the more common case in which thermodynamic forces are evaluated at the same instant of time and the EP rate remains strictly positive, see for instance Ref.~ \cite{Maes2000,Bauer2016,Brandner2016}.  


\section{Drude-Sommerfeld entropy production \label{sec.drudemodel}}

In the Drude model the charge carriers in an ideal metal are described as a classical free electron gas which obeys Maxwell-Boltzmann statistics \cite{Drude1900I,Drude1900II}. In its extension to the quantum domain, the equilibrium statistics is upgraded to the Fermi-Dirac one and the quantum effects are encoded in the density $N$ of charge carriers at the Fermi level, in the behavior of the relaxation time with temperature, and in the effective (band) mass $m$ of the electrons \cite{Sommerfeld1928,bardeen1940,Ashcroft1976}. 

Whereas an ideal gas is strictly free of collisions, assuming a finite relaxation time in the Drude model implies a finite mean free path.  This mean free path is justified by considering the scattering of the charge carriers with phonons and impurities \cite{bardeen1940}.  It can be phenomenologically introduced expressing the collision term of the Boltzmann equation as \cite{Ashcroft1976}, $\left(\partial \rho/\partial t\right)_{coll} = -\delta \rho/\tau_{R}\,$,  where $\tau_{R}$, $\rho$ and $\delta \rho = \rho-\rho_{0}$ denote the relaxation time, the non-equilibrium distribution and the deviation from the initial equilibrium distribution $\rho_{0}$, respectively. 

Hence, we obtain for the conductivity 
\begin{equation}
\sigma(\omega) = \frac{\sigma_{0}}{1-i\omega\,\tau_{R}}\,,
\label{eq:dsconduc}
\end{equation}
where $\sigma_{0}=N q^{2} \tau_{R}/m$ is the DC conductivity in the zero-frequency limit \citep{Ashcroft1976} and $q$ denotes the electric charge. 

Computing the inverse Fourier transform of Eq.~\eqref{eq:dsconduc}, we obtain the response function
\begin{equation}
\Phi(t) = \Theta(t)\, \frac{\sigma_{0}}{\tau_{R}}\e{-|t|/\tau_{R}}\,,
\label{eq:respds}
\end{equation}
and $\Theta(t)$ is the Heaviside step function. Hence, the EP rate \eqref{eq:eprdelay} becomes
\begin{equation}
\dot{\overline{\Sigma}} =\frac{\sigma_0}{T\tau_{R}}\, E(t) \int_{0}^{t}dt'\,\e{-|t-t'|/\tau_{R}}\, E(t')\,.
\label{eq:eprDS}
\end{equation}

\begin{figure}
\centering
\includegraphics[width=0.45\textwidth]{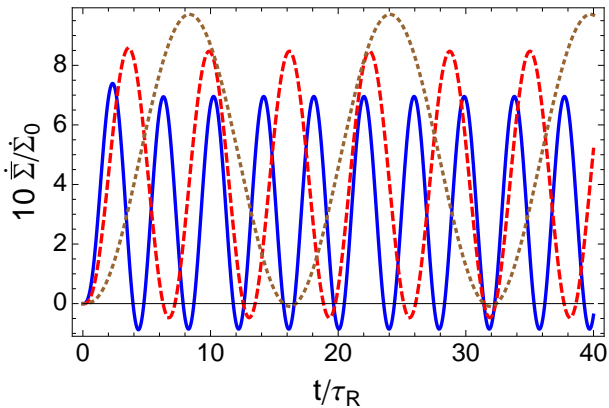}
\caption{(Color online) Entropy production rate (\ref{eq:eprharm}) as a function of $t/\tau_{R}$ for different values of $\omega_{0}\tau_{R}$. Solid-blue, dashed-red and dotted-brown lines correspond to $\omega_{0}\tau_{R} = 0.8$, $0.5$ and $0.2$, respectively. Units are $\dot{\Sigma}_0 \equiv \sigma_0 E_0^2/T$. }
\label{fig:eprDS}
\end{figure}

Assuming a monochromatic electric field, $E(t)=E_0 \sin\left(\omega_0 t\right)$, and employing Eq.~\eqref{eq:eprDS}, the entropy production rate \eqref{eq:eprdelay} reads,
\begin{equation}
\begin{split}
&\dot{\overline{\Sigma}} = \frac{1}{T}\frac{\sigma_0\,E^{2}_{0}\,\sin{(\omega_{0} t)}}{1+(\omega_{0}\tau_{R})^{2}}\\
&\times \left[\omega_{0}\tau_{R}\,\e{-t/\tau_{R}}-\omega_{0}\tau_{R}\,\cos{(\omega_{0} t)}+\sin{(\omega_{0} t)}\right].
\end{split}
\label{eq:eprharm}
\end{equation}
Figure~\ref{fig:eprDS} illustrates Eq.~\eqref{eq:eprharm} for different values of $\omega_{0}\tau_{R}$. As discussed in Ref.~\cite{Bonanca2021}, negative values of $\dot{\overline{\Sigma}}$ persist within a small vicinity of $\omega_{0} t = n\pi$, $n=1, 2, 3,\ldots$, as $\omega_{0} \tau_{R}$ decreases and vanish only in the limit $\omega_{0}\tau_{R}\to 0$.

The emergence of negative values of $\dot{\overline{\Sigma}}$ can be understood qualitatively from Eq.~(\ref{eq:eprDS}).  $\dot{\overline{\Sigma}}$ is given by the product of a time-dependent electric field and a convolution between the same field (evaluated at a previous time) and the response function. This describes a delay between the response of the system and the electric field. Thus, if the function $E(t)$ is non-monotonic and acquires negative values,  negative values of $\dot{\overline{\Sigma}}$ occur \footnote{Note, however, that it has been shown that the entropy production, $\overline{\Sigma}=\int_{0}^{t}dt'\,\dot{\overline{\Sigma}} $, itself remains positive at all times \cite{Bonanca2020a,Bonanca2021}. }.

\section{Fluctuation theorem \label{sec.ft}}

As mentioned before, Drude originally considered the charge carriers as a classical gas of particles following Maxwell-Boltzmann statistics. Although this leads to wrong predictions of the DC conductivity $\sigma_{0}$, it does not change the form of $\sigma(\omega)$ given by Eq.~(\ref{eq:dsconduc}). This means that the qualitative features of the EP rate $\dot{\overline{\Sigma}}$ and its expression (\ref{eq:eprharm}) remain unaltered in the original Drude model, although with a different value of $\sigma_{0}$.

For the next part of the analysis, we restrict ourselves to the classical Drude model and develop the corresponding stochastic thermodynamics. Here, $N$ denotes the \emph{number} (instead of density) of non-interacting charge carriers and by $J_{e}$ we denote the electrical current (instead of current density). We start by combining Ohm's law \eqref{eq:ohmslaw} with the Drude conductivity \eqref{eq:dsconduc},
\begin{equation}
\hat{J}_{e}(\omega) = \frac{\sigma_{0}}{1-i\omega\tau_{R}} \hat{E}(\omega)\,,
\label{eq:ohmslaw2}
\end{equation}
which in time domain takes the form of the following equation of motion for $J_{e}(t)$ ,
\begin{equation}
\frac{dJ_{e}}{dt} + \frac{J_{e}}{\tau_{R}} = \frac{\sigma_{0}}{\tau_{R}}E(t)\,.
\label{eq.currenteq}
\end{equation}
Recalling that $J_{e}(t)$ is defined as $ J_{e}(t) = N q \langle p(t)\rangle/m$,  where $\langle p(t)\rangle$ is the average momentum of a charge carrier along the direction of the electric field, Eq.~(\ref{eq.currenteq}) yields
\begin{equation}
\frac{d\langle p(t)\rangle}{dt} + \frac{\langle p(t)\rangle}{\tau_{R}} = q E(t)\,,
\label{eq.averagemomentum}
\end{equation}
and $\sigma_{0} = N q^{2}\tau_{R}/m$.

Fluctuations can be introduced by removing the average in Eq.~(\ref{eq.averagemomentum}) and introducing a Gaussian-distributed white noise $f_{i}(t)$ with zero mean acting on the $i$th charged particle. The equation of motion for the fluctuating linear momentum of the $i$th charge along the direction of $\mathbf{E}$(t) then reads
\begin{equation}
\frac{d p_{i}(t)}{dt} + \frac{p_{i}(t)}{\tau_{R}} = q E(t) + f_{i}(t)\,,
\label{eq.brownianm}
\end{equation}
which is nothing but Brownian motion under the external force $q E(t)$.  The fluctuating current $j_{e}(t)$ then is,
\begin{equation}
j_{e}(t) = \frac{q}{m}\sum_{i=1}^{N} p_{i}(t)\,,
\end{equation}
and we define the fluctuating entropy production,
\begin{equation}
\Sigma(t) = \frac{1}{T}\int_{0}^{t}dt'\,j_{e}(t') E(t')\,,
\label{eq.stochasticep}
\end{equation}
inspired by irreversible thermodynamics (see Eq.~\eqref{eq.epit}).

Since the $p_{i}(t)$ are Gaussian distributed and $\Sigma(t)$ is a linear transformation of their sum, the probability distribution of $\Sigma(t)$ will be Gaussian, as well.  See Refs.~\cite{zon2003,aquino2010} for similar arguments. From the solution of Eq.~(\ref{eq.brownianm}) and expression (\ref{eq.stochasticep}), it can be shown that (see App.~\ref{app.mom})
\begin{equation}
\langle \Sigma(t) \rangle = \frac{N q^{2}}{m T} \int_{0}^{t}dt' \int_{0}^{t'}dt''\,E(t')\,e^{-(t'-t'')/\tau_{R}}\,E(t'')\,.
\label{eq.averageep}
\end{equation}
when the linear momenta of the charge carriers are distributed according to Maxwell's distribution at $t=0$. 

Under the same assumptions, we have (see App.~\ref{app.mom})
\begin{eqnarray}
\lefteqn{\langle \Sigma^{2}(t)\rangle = \langle \Sigma(t)\rangle^{2}} \nonumber\\
&+& k_{B}\frac{N q^{2}}{m T}\int_{0}^{t}dt' \int_{0}^{t}dt''\,E(t')\,e^{-|t'-t''|/\tau_{R}}\,E(t'')\,,
\end{eqnarray}
and we obtain
\begin{equation}
\mathrm{Var}\left[\Sigma(t)\right] = \langle \Sigma^{2}(t)\rangle - \langle \Sigma(t)\rangle^{2}= 2 k_{B} \langle \Sigma(t)\rangle\,.
\end{equation}
The probability distribution of $\Sigma$ then reads
\begin{equation}
P(\Sigma) = \frac{1}{\sqrt{2\pi\mathrm{Var}\left[\Sigma\right]}} \exp{\left(-\frac{\left[ \Sigma - \langle \Sigma\rangle\right]^{2}}{2\mathrm{Var}\left[\Sigma\right]}\right)}\,,
\end{equation}
which implies the detailed Fluctuation Theorem,
\begin{equation}
\frac{P(\Sigma)}{P(-\Sigma)} = \exp{\left[ \frac{2\langle\Sigma\rangle \Sigma}{\mathrm{Var}\left(\Sigma\right)}\right]} = e^{\Sigma/k_{B}}\,.
\label{eq.ft}
\end{equation}
and its integral version, $\left\langle \e{-\Sigma/k_{B}}\right\rangle = 1$

Note that the average EP rate is,
\begin{equation}
\label{eq:EPR}
\langle \dot{\Sigma}(t)\rangle = \frac{Nq}{mT}\langle p(t)\rangle E(t)\\
\end{equation}
whose time integral is identical to Eq.~(\ref{eq.averageep}),  and with different values and units of $\sigma_{0}$ in Eq.~(\ref{eq:eprDS}).  Hence, we conclude that negative values of the EP rate are not incompatible with the Fluctuation Theorem (\ref{eq.ft}) and, hence, the second law. 

\section{Entropy production from microreversibility \label{sec.pathint}}

As a next step, we show that the expression (\ref{eq.stochasticep}) for the EP appears naturally from microreversibility \cite{Crooks1999}, i.e., from the comparison of probabilities of observing a given trajectory and its time-reversed conjugated twin \cite{Chernyak2006,Imparato2006,Deffner2011EPL,Pal2020NJP}.

Considering Eq.~(\ref{eq.brownianm}) for a single charge carrier, its corresponding Fokker-Planck equation reads \cite{Risken} (we have dropped the index $i$)
\begin{eqnarray}
\frac{\partial}{\partial t}\mathcal{W}(p,t) &=& -\frac{\partial}{\partial p}\left[ \left( -\frac{p}{\tau_{R}}+q E(t)\right) \mathcal{W}(p,t)\right] \nonumber\\
&+& \frac{\partial^{2}}{\partial p^{2}}\left( \frac{m k_{B}T}{\tau_{R}} \mathcal{W}(p,t)\right)\,,
\label{eq.fp}
\end{eqnarray}
where $\mathcal{W}(p,t)$ denotes the probability density of linear momentum at a given $t$. Following the standard procedures \cite{Risken}, it is straightforward to obtain the path-integral solution of Eq.~(\ref{eq.fp}) \cite{Chernyak2006,Imparato2006,Risken}.

In this formulation, the time evolution of $\mathcal{W}(p,t)$ is expressed in terms of probabilities of trajectories. Denoting by $\mathcal{P}[\Gamma|p_{0}]$ the conditional probability of observing a trajectory $\Gamma$ that starts at $p_{0}$, its expression reads
\begin{eqnarray}
\mathcal{P}[\Gamma|p_{0}] = \mathcal{N} \exp{\bigg\{ -\int_{0}^{t}dt'\, \mathcal{L}(p(t'),\dot{p}(t');E(t'))\bigg\}}\,,\nonumber\\
\label{eq.probtrajf}
\end{eqnarray}
where
\begin{equation}
\mathcal{L}(p(t'),\dot{p}(t');E(t')) = \frac{\left[ \dot{p}(t')-(-p(t')/\tau_{R} + q E(t'))\right]^{2}}{(4m k_{B}T/\tau_{R})}\,,
\label{eq.lagrange}
\end{equation}
and $\mathcal{N}$ is a normalization factor \cite{Risken}.

Following Ref.~\cite{Chernyak2006}, we define $\Gamma^{\dagger}$ as the time-reversed conjugated twin of $\Gamma$ for the time-reversed electric field $E^{R}(t)$. Thus, the probability of observing $\Gamma^{\dagger}$ is given by
\begin{eqnarray}
\mathcal{P}[\Gamma^{\dagger}|-p_{t}]&&=\nonumber\\
\mathcal{N} &&\exp{\bigg\{ -\int_{0}^{t}dt'\, \mathcal{L}(p^{\dagger}(t'),\dot{p}^{\dagger}(t');E^{R}(t'))\bigg\}}\,,\nonumber\\
\label{eq.probtrajr}
\end{eqnarray}
where the initial condition $-p_{t}$ of $\Gamma^{\dagger}$ is the time-reversal of the final point of $\Gamma$. 

Combining Eqs.~(\ref{eq.probtrajf}) and (\ref{eq.probtrajr}) we have the microreversibility condition (see App.~\ref{app.pathint}),
\begin{eqnarray}
\lefteqn{\frac{\mathcal{P}[\Gamma|p_{0}]}{\mathcal{P}[\Gamma^{\dagger}|-p_{t}]} =}\nonumber\\
&& \exp{\bigg\{-\frac{(p_{t}^{2}-p_{0}^{2})}{2mk_{B}T}+\frac{q}{mk_{B}T}\int_{0}^{t}dt'\,p(t')E(t')\bigg\}}\,,
\end{eqnarray}
where the second term in the exponent  is proportional to the single-particle version of Eq.~(\ref{eq.stochasticep}). This term can be understood as proportional to the fluctuating power input due to the electric field. On the other hand, the first term in the exponent is proportional to the variation of the stochastic internal energy (which in the present case is purely kinetic) between the final and initial states. Hence, energy conservation implies that the entire exponent is equal to the heat absorbed by the heat bath in units of $k_{B}T$ \cite{Chernyak2006,Imparato2006,Crooks1999,Pal2020NJP}. 

\section{Entropy production from Shannon information \label{sec.infoep}}

In the preceding sections, we showed how the fluctuating entropy production can be found in full consistency with linear irreversible thermodynamics. In stochastic thermodynamics, however, the EP is often obtained from the Shannon information \cite{peliti2021},
\begin{equation}
S(t) = -k_{B}\int dp\,\mathcal{W}(p,t) \ln{\mathcal{W}(p,t)}\,,
\label{eq.shannon}
\end{equation}
for the solution $\mathcal{W}(p,t)$ of Eq.~(\ref{eq.fp}) \cite{seifert2005}. Following standard arguments \cite{Tome2010,peliti2021}, we have
\begin{eqnarray}
\frac{1}{k_{B}}\frac{dS}{dt} 
&=&\frac{\tau_{R}}{m k_{B} T}\int dp\, \frac{\mathcal{J}(p,t)^{2}}{\mathcal{W}(p,t)} + \frac{1}{m k_{B} T}\int dp\, p \mathcal{J}(p,t)	\nonumber\\
& = & \Pi(t)-\Omega(t)\,,
\label{eq.infoeprate}
\end{eqnarray}
where
\begin{eqnarray}
\mathcal{J}(p,t) = -\frac{p}{\tau_{R}}\mathcal{W}(p,t)-\frac{\partial}{\partial p}\left( \frac{m k_{B} T}{\tau_{R}}\mathcal{W}(p,t)\right)\,.
\end{eqnarray}
The term denoted by $\Omega(t)$,
\begin{equation}
\Omega(t) = -\frac{1}{m k_{B} T}\int dp\, p \mathcal{J}(p,t)\,,
\end{equation}
is typically called the entropy flux  \cite{peliti2021} and {the non-negative term,
\begin{equation}
\Pi(t) = \frac{\tau_{R}}{m k_{B} T}\int dp\, \mathcal{W}(p,t)\left(\frac{\mathcal{J}(p,t)}{\mathcal{W}(p,t)}\right)^{2}\,,
\label{eq.infoeprate2}
\end{equation}
is taken as the EP rate \cite{Tome2010,Esposito2010c,peliti2021}.

The terminology and interpretation of the terms in Eq.~(\ref{eq.infoeprate}) make contact with the balance equation (\ref{eq.balanceEP}) introduced in the macroscopic approach of irreversible thermodynamics. However, the fact that $\Pi(t)$ is always non-negative already shows a strong disagreement with the EP rates discussed in the previous sections. Additionally, it is easy to see that $\Pi(t)$ does not have the bilinear form discussed in Sec.~\ref{sec.irrthermo}. 

Consider the solution of Eq.~(\ref{eq.fp}) \cite{aquino2010,ferrari2003},
\begin{eqnarray}
\mathcal{W}(p,t)=\left(2\pi m k_{B} T\right)^{-1/2} \exp{\left( -\frac{\left[ p - \langle p(t)\rangle\right]^{2}}{2 m k_{B} T}\right)}\,,
\label{eq.solution}
\end{eqnarray}
for an initial Maxwell-Boltzmann distribution. Then, the average momentum $\langle p(t)\rangle$ is,
\begin{equation}
\langle p(t)\rangle = \int_{0}^{t}dt' e^{-(t-t')/\tau_{R}} q E(t')\,,
\end{equation}
which is obtained from Eq.~(\ref{eq.averagemomentum}). Plugging expression~(\ref{eq.solution}) in Eq.~\eqref{eq.infoeprate2}, it is straightforward to show that,
\begin{equation}
\label{eq:EPRstoch}
k_{B}\Pi(t) = \frac{\langle p(t)\rangle^{2}}{m T\tau_{R}}\,,
\end{equation}
which is clearly different from current, $q\langle p(t)\rangle/m$, times thermodynamic force, $E(t)/T$, which we had before in Eq.~\eqref{eq:EPR}. However, note that Eqs.~\eqref{eq:EPR} and \eqref{eq:EPRstoch} do become identical in the limit of infinitely slow driving, i.e, $\omega_{0}\tau_{R}\to 0$. Equation~(\ref{eq:ohmslaw2}) tells us that the delay between $J_{e}(t)$ and $E(t)$ vanishes in this limit, i.e., the current becomes simply proportional to the electric field.

Additionally, the solution (\ref{eq.solution}) leads to the following expression for the entropy flux $\Omega(t)$,
\begin{equation}
k_{B}\Omega(t) = \frac{\langle p(t)\rangle^{2}}{m T\tau_{R}}\,.
\end{equation}
Together with Eq.~(\ref{eq:EPRstoch}), this implies that the time derivative of the Shannon information (\ref{eq.shannon}) is zero whether the system is in the transient or long-time state.

\section{Concluding remarks}

Motivated by the macroscopic approach of irreversible thermodynamics, we have defined an expression for the fluctuating EP in the classical Drude model. We have shown that this quantity fulfills a fluctuation theorem despite the existence of negative values of its corresponding rate. Hence, this shows, for a paradigmatic example, that negative entropy production rates are not incompatible with the fluctuation theorem and the second law. Moreover, we have shown that the entropy production rate obtained from the Shannon entropy and the Fokker-Planck equation contrasts with our expression and does not have the bilinear form predicted by irreversible thermodynamics.
In addition, we have shown that the Shannon information remains constant along the transition from the initial equilibrium state to long-time nonequilibrium one and that the entropy production obtained from it is simply proportional to the power related to the friction force. Our expression, however, is a measure of the total amount of energy absorbed by the insulated system composed of the set of charge carriers and its heat bath.

\appendix
\section{Irreversible thermodynamics \label{app.irrtherm}}

In this appendix, we give a short derivation of Eq.~(\ref{eq.epit}) for the EP rate in irreversible thermodynamics. The hypothesis of local equilibrium allows to extend standard relations of equilibrium thermodynamics to the non-equilibrium regime \cite{reichl}.  For instance, the combination of the first with the second law,
\begin{equation}
T ds = du - \mu dn\,,
\label{eq.firstsecondlaws}
\end{equation}
and the Gibbs-Duhem equation,
\begin{equation}
s dT = -dP-nd\mu\,,
\label{eq.gibbsduhem}
\end{equation}
are assumed to hold when the temperature $T$, the pressure $P$ and the chemical potential $\mu$ along with the densities of entropy $s$, internal energy $u$ and particles $n$, are treated as scalar fields.

Using Eqs.~(\ref{eq.firstsecondlaws}) and (\ref{eq.gibbsduhem}), it is possible to write \cite{reichl}
\begin{equation}
\frac{du}{dt}-T\frac{ds}{dt}-\mu\frac{d\mu}{dt} = \frac{\partial u}{\partial t}-T\frac{\partial s}{\partial t}-\mu\frac{\partial\mu}{\partial t}=0\,,
\label{eq.balancerel}
\end{equation}
through which the balance equations can be connected. Considering Eqs.~(\ref{eq.balanceIE}), (\ref{eq.balanceEP}) and conservation of particles,
\begin{equation}
\frac{\partial n}{\partial t} + \nabla\cdot\mathbf{J}_{n} = 0\,,
\end{equation}
it is possible to find for the entropy flux,
\begin{equation}
\mathbf{J}_{s} = \frac{\mathbf{J}_{u}}{T}-\mu\frac{\mathbf{J}_{n}}{T}\,,
\end{equation}
and the EP rate,
\begin{equation}
\dot{\overline{\Sigma}} =\mathbf{J}_{u}\cdot\nabla\left( \frac{1}{T}\right) -\mathbf{J}_{n}\cdot\nabla\left( \frac{\mu}{T}\right)+ \frac{ \mathbf{J}_{e}\cdot \mathbf{E}}{T}\,.
\label{eq.eprateapp}
\end{equation}

Under the conditions of uniform temperature and pressure, the Gibbs-Duhem equation in the form,
\begin{equation}
\nabla P + s\nabla T + n\nabla\mu=0\,,
\end{equation}
implies that $\nabla\mu = 0$ and Eq.~(\ref{eq.eprateapp}) reduces to
\begin{equation}
\dot{\overline{\Sigma}} = \frac{ \mathbf{J}_{e}\cdot \mathbf{E}}{T}\,.
\end{equation}

\section{The first two moments of $\Sigma(t)$ \label{app.mom}}

To calculate the first moment $\langle \Sigma(t)\rangle$, we start with Eq.~(\ref{eq.stochasticep}) and the solution $p(t)$ of Eq.~(\ref{eq.brownianm}),
\begin{eqnarray}
p_{i}(t) = p_{i}(0)e^{-t/\tau_{R}} &+&\int_{0}^{t}dt' e^{-(t-t')/\tau_{R}} q E(t') \nonumber\\
&+& \int_{0}^{t}dt' e^{-(t-t')/\tau_{R}} f_{i}(t')\,.
\label{eq.sollangevin}
\end{eqnarray}
Assuming that the system is in equilibrium at $t=0$, we insert the previous expression in Eq.~(\ref{eq.stochasticep}) and, after taking the average, Eq.~(\ref{eq.averageep}) is obtained using that $\langle p_{i}(0)\rangle = 0$ and $\langle f_{i}(t)\rangle=0$.

To obtain the second moment $\langle \Sigma^{2}(t)\rangle$, we take the average of Eq.~(\ref{eq.stochasticep}) squared, which reads
\begin{eqnarray}
\langle \Sigma^{2}(t)\rangle = \left( \frac{q}{m T}\right)^{2}\int_{0}^{t}ds \int_{0}^{t}du E(s) E(u) \sum_{i,j}\langle p_{i}(s) p_{j}(u)\rangle\,.
\nonumber\\
\label{eq.secondmoment}
\end{eqnarray}
Using again the solution (\ref{eq.sollangevin}), the initial condition $\langle p_{i}(0)p_{j}(0)\rangle = \delta_{ij} m k_{B}T$ and the white-noise property
\begin{equation}
\langle f_{i}(t)f_{j}(t')\rangle = \frac{2mk_{B}T}{\tau_{R}} \delta_{ij}\delta(t-t')\,,
\end{equation}
where $\delta_{ij}$ is Kronecker's delta and $\delta(t)$ is Dirac's delta function, it is possible to show that
\begin{eqnarray}
\lefteqn{\langle p_{i}(s)p_{j}(u)\rangle = \delta_{ij} m k_{B} T e^{-|s-u|/\tau_{R}} }\nonumber\\
&&+ q^{2}\int_{0}^{s}dt'\int_{0}^{u}dt''\,e^{-(s+u-t'-t'')/\tau_{R}} E(t')E(t'') .
\end{eqnarray}

Plugging the previous expression into Eq.~(\ref{eq.secondmoment}), we finally obtain
\begin{eqnarray}
\langle\Sigma^{2}(t)\rangle &=& \langle\Sigma(t)\rangle^{2} \nonumber\\
&+& k_{B}\frac{N q^{2}}{m T} \int_{0}^{t}ds \int_{0}^{t}du\,e^{-|s-u|/\tau_{R}} E(s) E(u)\nonumber\\
&=& \langle\Sigma(t)\rangle^{2} + 2 k_{B} \langle\Sigma(t)\rangle\,.
\end{eqnarray}

\section{Conditional probability of time-reversed trajectories \label{app.pathint}}

In this appendix, we explain what is meant by the ``action" appearing in Eq.~(\ref{eq.probtrajr}). Firstly, we clarify that $p^{\dagger}(t')=-p(t-t')$, $\dot{p}^{\dagger}(t')=\dot{p}(t-t')$ and $E^{R}(t') = E(t-t')$, where $p(t)$ refers to the trajectory $\Gamma$ whose evolution is subjected to $E(t)$. Using then Eq.~(\ref{eq.lagrange}), we obtain
\begin{eqnarray}
\lefteqn{\mathcal{L}(p^{\dagger}(t'),\dot{p}^{\dagger}(t');E^{R}(t')) =}\nonumber\\
&&\qquad\; \frac{\left[ \dot{p}(t-t')-(p(t-t')/\tau_{R} + q E(t-t'))\right]^{2}}{(4m k_{B}T/\tau_{R})}\,.
\end{eqnarray}
which implies that,
\begin{eqnarray}
\lefteqn{\int_{0}^{t}dt'\mathcal{L}(p^{\dagger}(t'),\dot{p}^{\dagger}(t');E^{R}(t')) =}\nonumber\\
&&\qquad\; \int_{0}^{t}ds'\,\frac{\left[ \dot{p}(s')-(p(s')/\tau_{R} + q E(s'))\right]^{2}}{(4m k_{B}T/\tau_{R})}\,,
\end{eqnarray}
after the change of variables $s'=t-t'$. Finally, the conditional probability of observing the trajectory $\Gamma^{\dagger}$ under the time-reversed field $E^{R}(t)$ reads
\begin{eqnarray}
\mathcal{P}[\Gamma^{\dagger}|-p_{t}]&&=\nonumber\\
\mathcal{N} &&\exp{\bigg\{ -\int_{0}^{t}ds'\, \frac{\left[ \dot{p}(s')-(p(s')/\tau_{R} + q E(s'))\right]^{2}}{(4m k_{B}T/\tau_{R})}\bigg\}}\,,\nonumber\\
\end{eqnarray}

\acknowledgements{The authors acknowledge comments by J. Thingna and K. Ptaszy\'nski. M. V. S. Bonan\c{c}a acknowledges financial support from FAPESP (Funda\c{c}\~{a}o de Amparo \`a Pesquisa do Estado de S\~ao Paulo) (Brazil) (Grant No. 2020/02170-4).}


\begin{thebibliography}{52}%
\makeatletter
\providecommand \@ifxundefined [1]{%
 \@ifx{#1\undefined}
}%
\providecommand \@ifnum [1]{%
 \ifnum #1\expandafter \@firstoftwo
 \else \expandafter \@secondoftwo
 \fi
}%
\providecommand \@ifx [1]{%
 \ifx #1\expandafter \@firstoftwo
 \else \expandafter \@secondoftwo
 \fi
}%
\providecommand \natexlab [1]{#1}%
\providecommand \enquote  [1]{``#1''}%
\providecommand \bibnamefont  [1]{#1}%
\providecommand \bibfnamefont [1]{#1}%
\providecommand \citenamefont [1]{#1}%
\providecommand \href@noop [0]{\@secondoftwo}%
\providecommand \href [0]{\begingroup \@sanitize@url \@href}%
\providecommand \@href[1]{\@@startlink{#1}\@@href}%
\providecommand \@@href[1]{\endgroup#1\@@endlink}%
\providecommand \@sanitize@url [0]{\catcode `\\12\catcode `\$12\catcode
  `\&12\catcode `\#12\catcode `\^12\catcode `\_12\catcode `\%12\relax}%
\providecommand \@@startlink[1]{}%
\providecommand \@@endlink[0]{}%
\providecommand \url  [0]{\begingroup\@sanitize@url \@url }%
\providecommand \@url [1]{\endgroup\@href {#1}{\urlprefix }}%
\providecommand \urlprefix  [0]{URL }%
\providecommand \Eprint [0]{\href }%
\providecommand \doibase [0]{http://dx.doi.org/}%
\providecommand \selectlanguage [0]{\@gobble}%
\providecommand \bibinfo  [0]{\@secondoftwo}%
\providecommand \bibfield  [0]{\@secondoftwo}%
\providecommand \translation [1]{[#1]}%
\providecommand \BibitemOpen [0]{}%
\providecommand \bibitemStop [0]{}%
\providecommand \bibitemNoStop [0]{.\EOS\space}%
\providecommand \EOS [0]{\spacefactor3000\relax}%
\providecommand \BibitemShut  [1]{\csname bibitem#1\endcsname}%
\let\auto@bib@innerbib\@empty
\bibitem [{\citenamefont {Callen}(1985)}]{Callen1985}%
  \BibitemOpen
  \bibfield  {author} {\bibinfo {author} {\bibfnamefont {H.}~\bibnamefont
  {Callen}},\ }\href
  {https://www.wiley.com/en-us/Thermodynamics+and+an+Introduction+to+Thermostatistics%2C+2nd+Edition-p-9780471862567}
  {\emph {\bibinfo {title} {Thermodynamics and an Introduction to
  Thermostastistics}}}\ (\bibinfo  {publisher} {Wiley},\ \bibinfo {address}
  {New York, USA},\ \bibinfo {year} {1985})\BibitemShut {NoStop}%
\bibitem [{\citenamefont {Prigogine}(1961)}]{prigogine1961}%
  \BibitemOpen
  \bibfield  {author} {\bibinfo {author} {\bibfnamefont {I.}~\bibnamefont
  {Prigogine}},\ }\href {https://books.google.com.br/books?id=wXcLtQEACAAJ}
  {\emph {\bibinfo {title} {Introduction to Thermodynamics of Irreversible
  Processes}}}\ (\bibinfo  {publisher} {Interscience},\ \bibinfo {address} {New
  York},\ \bibinfo {year} {1961})\BibitemShut {NoStop}%
\bibitem [{\citenamefont {De~Groot}\ and\ \citenamefont
  {Mazur}(2013)}]{mazur2011}%
  \BibitemOpen
  \bibfield  {author} {\bibinfo {author} {\bibfnamefont {S.~R.}\ \bibnamefont
  {De~Groot}}\ and\ \bibinfo {author} {\bibfnamefont {P.}~\bibnamefont
  {Mazur}},\ }\href
  {https://books.google.com.br/books?id=mfFyG9jfaMYC&dq=de+groot+mazur&lr=&source=gbs_navlinks_s}
  {\emph {\bibinfo {title} {Non-equilibrium thermodynamics}}}\ (\bibinfo
  {publisher} {Courier Corporation},\ \bibinfo {address} {New York},\ \bibinfo
  {year} {2013})\BibitemShut {NoStop}%
\bibitem [{\citenamefont {Zwanzig}(2001)}]{zwanzig2001}%
  \BibitemOpen
  \bibfield  {author} {\bibinfo {author} {\bibfnamefont {Robert}\ \bibnamefont
  {Zwanzig}},\ }\href
  {https://www.google.com/books/edition/Nonequilibrium_Statistical_Mechanics/4cI5136OdoMC?hl=en&gbpv=1&printsec=frontcover}
  {\emph {\bibinfo {title} {Nonequilibrium statistical mechanics}}}\ (\bibinfo
  {publisher} {Oxford University Press},\ \bibinfo {address} {Oxford, UK},\
  \bibinfo {year} {2001})\BibitemShut {NoStop}%
\bibitem [{\citenamefont {Bonan\c{c}a}\ \emph {et~al.}(2021)\citenamefont
  {Bonan\c{c}a}, \citenamefont {Naz\'e},\ and\ \citenamefont
  {Deffner}}]{Bonanca2021}%
  \BibitemOpen
  \bibfield  {author} {\bibinfo {author} {\bibfnamefont {Marcus V.~S.}\
  \bibnamefont {Bonan\c{c}a}}, \bibinfo {author} {\bibfnamefont {Pierre}\
  \bibnamefont {Naz\'e}}, \ and\ \bibinfo {author} {\bibfnamefont {Sebastian}\
  \bibnamefont {Deffner}},\ }\bibfield  {title} {\enquote {\bibinfo {title}
  {Negative entropy production rates in drude-sommerfeld metals},}\ }\href
  {\doibase 10.1103/PhysRevE.103.012109} {\bibfield  {journal} {\bibinfo
  {journal} {Phys. Rev. E}\ }\textbf {\bibinfo {volume} {103}},\ \bibinfo
  {pages} {012109} (\bibinfo {year} {2021})}\BibitemShut {NoStop}%
\bibitem [{\citenamefont {Williams}\ \emph {et~al.}(2007)\citenamefont
  {Williams}, \citenamefont {Evans},\ and\ \citenamefont
  {Mittag}}]{Williams2007}%
  \BibitemOpen
  \bibfield  {author} {\bibinfo {author} {\bibfnamefont {Stephen~R.}\
  \bibnamefont {Williams}}, \bibinfo {author} {\bibfnamefont {Denis~J.}\
  \bibnamefont {Evans}}, \ and\ \bibinfo {author} {\bibfnamefont {Emil}\
  \bibnamefont {Mittag}},\ }\bibfield  {title} {\enquote {\bibinfo {title}
  {Negative entropy production in oscillatory processes},}\ }\href {\doibase
  https://doi.org/10.1016/j.crhy.2007.05.007} {\bibfield  {journal} {\bibinfo
  {journal} {C. R. Phys.}\ }\textbf {\bibinfo {volume} {8}},\ \bibinfo {pages}
  {620 -- 624} (\bibinfo {year} {2007})}\BibitemShut {NoStop}%
\bibitem [{\citenamefont {Thingna}\ \emph {et~al.}(2017)\citenamefont
  {Thingna}, \citenamefont {Barra},\ and\ \citenamefont
  {Esposito}}]{thingna2017}%
  \BibitemOpen
  \bibfield  {author} {\bibinfo {author} {\bibfnamefont {J.}~\bibnamefont
  {Thingna}}, \bibinfo {author} {\bibfnamefont {F.}~\bibnamefont {Barra}}, \
  and\ \bibinfo {author} {\bibfnamefont {M.}~\bibnamefont {Esposito}},\
  }\bibfield  {title} {\enquote {\bibinfo {title} {Kinetics and thermodynamics
  of a driven open quantum systems},}\ }\href {\doibase
  10.1103/PhysRevE.96.052132} {\bibfield  {journal} {\bibinfo  {journal} {Phys.
  Rev. E}\ }\textbf {\bibinfo {volume} {96}},\ \bibinfo {pages} {052132}
  (\bibinfo {year} {2017})}\BibitemShut {NoStop}%
\bibitem [{\citenamefont {Chakraborti}\ \emph {et~al.}(2021)\citenamefont
  {Chakraborti}, \citenamefont {Dhar}, \citenamefont {Goldstein}, \citenamefont
  {Kundu},\ and\ \citenamefont {Lebowitz}}]{chakraborti2021}%
  \BibitemOpen
  \bibfield  {author} {\bibinfo {author} {\bibfnamefont {Subhadip}\
  \bibnamefont {Chakraborti}}, \bibinfo {author} {\bibfnamefont {Abhishek}\
  \bibnamefont {Dhar}}, \bibinfo {author} {\bibfnamefont {Sheldon}\
  \bibnamefont {Goldstein}}, \bibinfo {author} {\bibfnamefont {Anupam}\
  \bibnamefont {Kundu}}, \ and\ \bibinfo {author} {\bibfnamefont {Joel~L.}\
  \bibnamefont {Lebowitz}},\ }\bibfield  {title} {\enquote {\bibinfo {title}
  {Entropy growth during free expansion of an ideal gas},}\ }\href
  {https://arxiv.org/abs/2109.07742} {\bibfield  {journal} {\bibinfo  {journal}
  {arXiv preprint arXiv:2109.07742}\ } (\bibinfo {year} {2021})}\BibitemShut
  {NoStop}%
\bibitem [{\citenamefont {Bhattacharya}\ \emph {et~al.}(2017)\citenamefont
  {Bhattacharya}, \citenamefont {Misra}, \citenamefont {Mukhopadhyay},\ and\
  \citenamefont {Pati}}]{bhattacharya2017}%
  \BibitemOpen
  \bibfield  {author} {\bibinfo {author} {\bibfnamefont {S.}~\bibnamefont
  {Bhattacharya}}, \bibinfo {author} {\bibfnamefont {A.}~\bibnamefont {Misra}},
  \bibinfo {author} {\bibfnamefont {C.}~\bibnamefont {Mukhopadhyay}}, \ and\
  \bibinfo {author} {\bibfnamefont {A.~K.}\ \bibnamefont {Pati}},\ }\bibfield
  {title} {\enquote {\bibinfo {title} {Exact master equation for a spin
  interacting with a spin bath: Non-markovianity and negative entropy
  production rate},}\ }\href
  {https://journals.aps.org/pra/abstract/10.1103/PhysRevA.95.012122} {\bibfield
   {journal} {\bibinfo  {journal} {Phys. Rev. A}\ }\textbf {\bibinfo {volume}
  {95}},\ \bibinfo {pages} {012122} (\bibinfo {year} {2017})}\BibitemShut
  {NoStop}%
\bibitem [{\citenamefont {Marcantoni}\ \emph {et~al.}(2017)\citenamefont
  {Marcantoni}, \citenamefont {Alipour}, \citenamefont {Benatti}, \citenamefont
  {Floreanini},\ and\ \citenamefont {Rezakhani}}]{marcantoni2017}%
  \BibitemOpen
  \bibfield  {author} {\bibinfo {author} {\bibfnamefont {S.}~\bibnamefont
  {Marcantoni}}, \bibinfo {author} {\bibfnamefont {S.}~\bibnamefont {Alipour}},
  \bibinfo {author} {\bibfnamefont {F.}~\bibnamefont {Benatti}}, \bibinfo
  {author} {\bibfnamefont {R.}~\bibnamefont {Floreanini}}, \ and\ \bibinfo
  {author} {\bibfnamefont {A.~T.}\ \bibnamefont {Rezakhani}},\ }\bibfield
  {title} {\enquote {\bibinfo {title} {Entropy production and non-markovian
  dynamical maps},}\ }\href
  {https://www.nature.com/articles/s41598-017-12595-x} {\bibfield  {journal}
  {\bibinfo  {journal} {Sci. Rep.}\ }\textbf {\bibinfo {volume} {7}},\ \bibinfo
  {pages} {12447} (\bibinfo {year} {2017})}\BibitemShut {NoStop}%
\bibitem [{\citenamefont {Xu}\ \emph {et~al.}(2018)\citenamefont {Xu},
  \citenamefont {Liu},\ and\ \citenamefont {Feng}}]{xu2018}%
  \BibitemOpen
  \bibfield  {author} {\bibinfo {author} {\bibfnamefont {Y.~Y.}\ \bibnamefont
  {Xu}}, \bibinfo {author} {\bibfnamefont {J.}~\bibnamefont {Liu}}, \ and\
  \bibinfo {author} {\bibfnamefont {M.}~\bibnamefont {Feng}},\ }\bibfield
  {title} {\enquote {\bibinfo {title} {Positive entropy production rate induced
  by non-markovianity},}\ }\href
  {https://journals.aps.org/pre/abstract/10.1103/PhysRevE.98.032102} {\bibfield
   {journal} {\bibinfo  {journal} {Phys. Rev. E}\ }\textbf {\bibinfo {volume}
  {98}},\ \bibinfo {pages} {032102} (\bibinfo {year} {2018})}\BibitemShut
  {NoStop}%
\bibitem [{\citenamefont {Popovic}\ \emph {et~al.}(2018)\citenamefont
  {Popovic}, \citenamefont {Vacchini},\ and\ \citenamefont
  {Campbell}}]{popovic2018}%
  \BibitemOpen
  \bibfield  {author} {\bibinfo {author} {\bibfnamefont {M.}~\bibnamefont
  {Popovic}}, \bibinfo {author} {\bibfnamefont {B.}~\bibnamefont {Vacchini}}, \
  and\ \bibinfo {author} {\bibfnamefont {S.}~\bibnamefont {Campbell}},\
  }\bibfield  {title} {\enquote {\bibinfo {title} {Entropy production and
  correlations in a controlled non-markovian setting},}\ }\href
  {https://journals.aps.org/pra/abstract/10.1103/PhysRevA.98.012130} {\bibfield
   {journal} {\bibinfo  {journal} {Phys. Rev. A}\ }\textbf {\bibinfo {volume}
  {98}},\ \bibinfo {pages} {012130} (\bibinfo {year} {2018})}\BibitemShut
  {NoStop}%
\bibitem [{\citenamefont {Strasberg}\ and\ \citenamefont
  {Esposito}(2019)}]{strasberg2019}%
  \BibitemOpen
  \bibfield  {author} {\bibinfo {author} {\bibfnamefont {P.}~\bibnamefont
  {Strasberg}}\ and\ \bibinfo {author} {\bibfnamefont {M.}~\bibnamefont
  {Esposito}},\ }\bibfield  {title} {\enquote {\bibinfo {title}
  {Non-markovianity and negative entropy production rates},}\ }\href
  {https://journals.aps.org/pre/abstract/10.1103/PhysRevE.99.012120} {\bibfield
   {journal} {\bibinfo  {journal} {Phys. Rev. E}\ }\textbf {\bibinfo {volume}
  {99}},\ \bibinfo {pages} {012120} (\bibinfo {year} {2019})}\BibitemShut
  {NoStop}%
\bibitem [{\citenamefont {Zwanzig}(1961)}]{zwanzig1961}%
  \BibitemOpen
  \bibfield  {author} {\bibinfo {author} {\bibfnamefont {R.}~\bibnamefont
  {Zwanzig}},\ }\bibfield  {title} {\enquote {\bibinfo {title} {Memory effects
  in irreversible thermodynamics},}\ }\href
  {https://doi.org/10.1103/PhysRev.124.983} {\bibfield  {journal} {\bibinfo
  {journal} {Phys. Rev.}\ }\textbf {\bibinfo {volume} {124}},\ \bibinfo {pages}
  {983} (\bibinfo {year} {1961})}\BibitemShut {NoStop}%
\bibitem [{\citenamefont {McLennan}(1964)}]{McLennan1964}%
  \BibitemOpen
  \bibfield  {author} {\bibinfo {author} {\bibfnamefont {J.~A.}\ \bibnamefont
  {McLennan}},\ }\bibfield  {title} {\enquote {\bibinfo {title} {Entropy
  production for a medium with memory},}\ }\href
  {https://doi.org/10.1063/1.1726025} {\bibfield  {journal} {\bibinfo
  {journal} {J. Chem. Phys.}\ }\textbf {\bibinfo {volume} {41}},\ \bibinfo
  {pages} {1159} (\bibinfo {year} {1964})}\BibitemShut {NoStop}%
\bibitem [{\citenamefont {Naz{\'e}}\ and\ \citenamefont
  {Bonan{\c{c}}a}(2020)}]{Bonanca2020a}%
  \BibitemOpen
  \bibfield  {author} {\bibinfo {author} {\bibfnamefont {P.}~\bibnamefont
  {Naz{\'e}}}\ and\ \bibinfo {author} {\bibfnamefont {M.~V.~S.}\ \bibnamefont
  {Bonan{\c{c}}a}},\ }\bibfield  {title} {\enquote {\bibinfo {title}
  {Compatibility of linear-response theory with the second law of
  thermodynamics and the emergence of negative entropy production rates},}\
  }\href {https://iopscience.iop.org/article/10.1088/1742-5468/ab54ba}
  {\bibfield  {journal} {\bibinfo  {journal} {J. Stat. Mech.: Theo. Exp.}\ ,\
  \bibinfo {pages} {013206}} (\bibinfo {year} {2020})}\BibitemShut {NoStop}%
\bibitem [{\citenamefont {Peliti}\ and\ \citenamefont
  {Pigolotti}(2021)}]{peliti2021}%
  \BibitemOpen
  \bibfield  {author} {\bibinfo {author} {\bibfnamefont {Luca}\ \bibnamefont
  {Peliti}}\ and\ \bibinfo {author} {\bibfnamefont {Simone}\ \bibnamefont
  {Pigolotti}},\ }\href
  {https://press.princeton.edu/books/hardcover/9780691201771/stochastic-thermodynamics}
  {\emph {\bibinfo {title} {Stochastic Thermodynamics}}}\ (\bibinfo
  {publisher} {Princeton University Press},\ \bibinfo {address} {Princeton, New
  Jersey, US},\ \bibinfo {year} {2021})\BibitemShut {NoStop}%
\bibitem [{\citenamefont {Seifert}(2005)}]{seifert2005}%
  \BibitemOpen
  \bibfield  {author} {\bibinfo {author} {\bibfnamefont {U.}~\bibnamefont
  {Seifert}},\ }\bibfield  {title} {\enquote {\bibinfo {title} {Entropy
  production along a stochastic trajectory and an integral fluctuation
  theorem},}\ }\href
  {https://journals.aps.org/prl/abstract/10.1103/PhysRevLett.95.040602}
  {\bibfield  {journal} {\bibinfo  {journal} {Phys. Rev. Lett.}\ }\textbf
  {\bibinfo {volume} {95}},\ \bibinfo {pages} {040602} (\bibinfo {year}
  {2005})}\BibitemShut {NoStop}%
\bibitem [{\citenamefont {Crooks}(1999)}]{Crooks1999}%
  \BibitemOpen
  \bibfield  {author} {\bibinfo {author} {\bibfnamefont {G.~E.}\ \bibnamefont
  {Crooks}},\ }\bibfield  {title} {\enquote {\bibinfo {title} {Entropy
  production fluctuation theorem and nonequilibrium work relation for free
  energy differences},}\ }\href {\doibase 10.1103/PhysRevE.60.2721} {\bibfield
  {journal} {\bibinfo  {journal} {Phys. Rev. E}\ }\textbf {\bibinfo {volume}
  {60}},\ \bibinfo {pages} {2721} (\bibinfo {year} {1999})}\BibitemShut
  {NoStop}%
\bibitem [{\citenamefont {Esposito}\ and\ \citenamefont {Van~den
  Broeck}(2010{\natexlab{a}})}]{Esposito2010a}%
  \BibitemOpen
  \bibfield  {author} {\bibinfo {author} {\bibfnamefont {M.}~\bibnamefont
  {Esposito}}\ and\ \bibinfo {author} {\bibfnamefont {C.}~\bibnamefont {Van~den
  Broeck}},\ }\bibfield  {title} {\enquote {\bibinfo {title} {Three detailed
  fluctuation theorems},}\ }\href
  {https://doi.org/10.1103/PhysRevLett.104.090601} {\bibfield  {journal}
  {\bibinfo  {journal} {Phys. Rev. Lett.}\ }\textbf {\bibinfo {volume} {104}},\
  \bibinfo {pages} {090601} (\bibinfo {year} {2010}{\natexlab{a}})}\BibitemShut
  {NoStop}%
\bibitem [{\citenamefont {Spohn}(1978)}]{spohn1978}%
  \BibitemOpen
  \bibfield  {author} {\bibinfo {author} {\bibfnamefont {H.}~\bibnamefont
  {Spohn}},\ }\bibfield  {title} {\enquote {\bibinfo {title} {Entropy
  production for quantum dynamical semigroups},}\ }\href
  {https://aip.scitation.org/doi/abs/10.1063/1.523789} {\bibfield  {journal}
  {\bibinfo  {journal} {J. Math. Phys.}\ }\textbf {\bibinfo {volume} {19}},\
  \bibinfo {pages} {1227} (\bibinfo {year} {1978})}\BibitemShut {NoStop}%
\bibitem [{\citenamefont {Prigogine}\ and\ \citenamefont
  {G{\'e}h{\'e}niau}(1986)}]{Prigogine1986}%
  \BibitemOpen
  \bibfield  {author} {\bibinfo {author} {\bibfnamefont {I.}~\bibnamefont
  {Prigogine}}\ and\ \bibinfo {author} {\bibfnamefont {J.}~\bibnamefont
  {G{\'e}h{\'e}niau}},\ }\bibfield  {title} {\enquote {\bibinfo {title}
  {Entropy, matter, and cosmology},}\ }\href {\doibase 10.1073/pnas.83.17.6245}
  {\bibfield  {journal} {\bibinfo  {journal} {PNAS}\ }\textbf {\bibinfo
  {volume} {83}},\ \bibinfo {pages} {6245--6249} (\bibinfo {year}
  {1986})}\BibitemShut {NoStop}%
\bibitem [{\citenamefont {Ruelle}(1996)}]{ruelle1996}%
  \BibitemOpen
  \bibfield  {author} {\bibinfo {author} {\bibfnamefont {D.}~\bibnamefont
  {Ruelle}},\ }\bibfield  {title} {\enquote {\bibinfo {title} {Positivity of
  entropy production in nonequilibrium statistical mechanics},}\ }\href
  {https://link.springer.com/article/10.1007/BF02175553} {\bibfield  {journal}
  {\bibinfo  {journal} {J. Stat. Phys.}\ }\textbf {\bibinfo {volume} {85}},\
  \bibinfo {pages} {1} (\bibinfo {year} {1996})}\BibitemShut {NoStop}%
\bibitem [{\citenamefont {Maes}\ \emph {et~al.}(2000)\citenamefont {Maes},
  \citenamefont {Redig},\ and\ \citenamefont {Moffaert}}]{Maes2000}%
  \BibitemOpen
  \bibfield  {author} {\bibinfo {author} {\bibfnamefont {Christian}\
  \bibnamefont {Maes}}, \bibinfo {author} {\bibfnamefont {Frank}\ \bibnamefont
  {Redig}}, \ and\ \bibinfo {author} {\bibfnamefont {Annelies~Van}\
  \bibnamefont {Moffaert}},\ }\bibfield  {title} {\enquote {\bibinfo {title}
  {On the definition of entropy production, via examples},}\ }\href {\doibase
  10.1063/1.533195} {\bibfield  {journal} {\bibinfo  {journal} {J. Math.
  Phys.}\ }\textbf {\bibinfo {volume} {41}},\ \bibinfo {pages} {1528--1554}
  (\bibinfo {year} {2000})}\BibitemShut {NoStop}%
\bibitem [{\citenamefont {Belandria}(2005)}]{Belandria2005}%
  \BibitemOpen
  \bibfield  {author} {\bibinfo {author} {\bibfnamefont {J.~I}\ \bibnamefont
  {Belandria}},\ }\bibfield  {title} {\enquote {\bibinfo {title} {Positive and
  negative entropy production in an ideal-gas expansion},}\ }\href {\doibase
  10.1209/epl/i2004-10508-7} {\bibfield  {journal} {\bibinfo  {journal}
  {Europhys. Lett. ({EPL})}\ }\textbf {\bibinfo {volume} {70}},\ \bibinfo
  {pages} {446--451} (\bibinfo {year} {2005})}\BibitemShut {NoStop}%
\bibitem [{\citenamefont {Esposito}\ and\ \citenamefont {Van~den
  Broeck}(2010{\natexlab{b}})}]{Esposito2010b}%
  \BibitemOpen
  \bibfield  {author} {\bibinfo {author} {\bibfnamefont {M.}~\bibnamefont
  {Esposito}}\ and\ \bibinfo {author} {\bibfnamefont {C.}~\bibnamefont {Van~den
  Broeck}},\ }\bibfield  {title} {\enquote {\bibinfo {title} {Three faces of
  the second law. i. master equation formulation},}\ }\href
  {https://doi.org/10.1103/PhysRevE.82.011143} {\bibfield  {journal} {\bibinfo
  {journal} {Phys. Rev. E}\ }\textbf {\bibinfo {volume} {82}},\ \bibinfo
  {pages} {011143} (\bibinfo {year} {2010}{\natexlab{b}})}\BibitemShut
  {NoStop}%
\bibitem [{\citenamefont {Van~den Broeck}\ and\ \citenamefont
  {Esposito}(2010)}]{Esposito2010c}%
  \BibitemOpen
  \bibfield  {author} {\bibinfo {author} {\bibfnamefont {C.}~\bibnamefont
  {Van~den Broeck}}\ and\ \bibinfo {author} {\bibfnamefont {M.}~\bibnamefont
  {Esposito}},\ }\bibfield  {title} {\enquote {\bibinfo {title} {Three faces of
  the second law. ii. fokker-planck formulation},}\ }\href
  {https://doi.org/10.1103/PhysRevE.82.011144} {\bibfield  {journal} {\bibinfo
  {journal} {Phys. Rev. E}\ }\textbf {\bibinfo {volume} {82}},\ \bibinfo
  {pages} {011144} (\bibinfo {year} {2010})}\BibitemShut {NoStop}%
\bibitem [{\citenamefont {Tom\'e}\ and\ \citenamefont
  {Oliveira}(2010)}]{Tome2010}%
  \BibitemOpen
  \bibfield  {author} {\bibinfo {author} {\bibfnamefont {T.}~\bibnamefont
  {Tom\'e}}\ and\ \bibinfo {author} {\bibfnamefont {M.~J.~de}\ \bibnamefont
  {Oliveira}},\ }\bibfield  {title} {\enquote {\bibinfo {title} {Entropy
  production in irreversible systems described by a fokker-planck equation},}\
  }\href {\doibase 10.1103/PhysRevE.82.021120} {\bibfield  {journal} {\bibinfo
  {journal} {Phys. Rev. E}\ }\textbf {\bibinfo {volume} {82}},\ \bibinfo
  {pages} {021120} (\bibinfo {year} {2010})}\BibitemShut {NoStop}%
\bibitem [{\citenamefont {Bauer}\ \emph {et~al.}(2016)\citenamefont {Bauer},
  \citenamefont {Brandner},\ and\ \citenamefont {Seifert}}]{Bauer2016}%
  \BibitemOpen
  \bibfield  {author} {\bibinfo {author} {\bibfnamefont {Michael}\ \bibnamefont
  {Bauer}}, \bibinfo {author} {\bibfnamefont {Kay}\ \bibnamefont {Brandner}}, \
  and\ \bibinfo {author} {\bibfnamefont {Udo}\ \bibnamefont {Seifert}},\
  }\bibfield  {title} {\enquote {\bibinfo {title} {Optimal performance of
  periodically driven, stochastic heat engines under limited control},}\ }\href
  {\doibase 10.1103/PhysRevE.93.042112} {\bibfield  {journal} {\bibinfo
  {journal} {Phys. Rev. E}\ }\textbf {\bibinfo {volume} {93}},\ \bibinfo
  {pages} {042112} (\bibinfo {year} {2016})}\BibitemShut {NoStop}%
\bibitem [{\citenamefont {Brandner}\ and\ \citenamefont
  {Seifert}(2016)}]{Brandner2016}%
  \BibitemOpen
  \bibfield  {author} {\bibinfo {author} {\bibfnamefont {Kay}\ \bibnamefont
  {Brandner}}\ and\ \bibinfo {author} {\bibfnamefont {Udo}\ \bibnamefont
  {Seifert}},\ }\bibfield  {title} {\enquote {\bibinfo {title} {Periodic
  thermodynamics of open quantum systems},}\ }\href {\doibase
  10.1103/PhysRevE.93.062134} {\bibfield  {journal} {\bibinfo  {journal} {Phys.
  Rev. E}\ }\textbf {\bibinfo {volume} {93}},\ \bibinfo {pages} {062134}
  (\bibinfo {year} {2016})}\BibitemShut {NoStop}%
\bibitem [{\citenamefont {Deffner}(2017)}]{deffner2017}%
  \BibitemOpen
  \bibfield  {author} {\bibinfo {author} {\bibfnamefont {S.}~\bibnamefont
  {Deffner}},\ }\bibfield  {title} {\enquote {\bibinfo {title} {Kibble-zurek
  scaling of the irreversible entropy production},}\ }\href
  {https://journals.aps.org/pre/abstract/10.1103/PhysRevE.96.052125} {\bibfield
   {journal} {\bibinfo  {journal} {Phys. Rev. E}\ }\textbf {\bibinfo {volume}
  {96}},\ \bibinfo {pages} {052125} (\bibinfo {year} {2017})}\BibitemShut
  {NoStop}%
\bibitem [{\citenamefont {Jabraoui}\ \emph {et~al.}(2020)\citenamefont
  {Jabraoui}, \citenamefont {Ouaskit}, \citenamefont {Richard},\ and\
  \citenamefont {Garden}}]{jabraoui2019}%
  \BibitemOpen
  \bibfield  {author} {\bibinfo {author} {\bibfnamefont {H}~\bibnamefont
  {Jabraoui}}, \bibinfo {author} {\bibfnamefont {S}~\bibnamefont {Ouaskit}},
  \bibinfo {author} {\bibfnamefont {J}~\bibnamefont {Richard}}, \ and\ \bibinfo
  {author} {\bibfnamefont {J-L}\ \bibnamefont {Garden}},\ }\bibfield  {title}
  {\enquote {\bibinfo {title} {Determination of the entropy production during
  glass transition: theory and experiment},}\ }\href
  {https://www.sciencedirect.com/science/article/abs/pii/S0022309320300247}
  {\bibfield  {journal} {\bibinfo  {journal} {J. Non-Cryst. Sol.}\ }\textbf
  {\bibinfo {volume} {533}},\ \bibinfo {pages} {119907} (\bibinfo {year}
  {2020})}\BibitemShut {NoStop}%
\bibitem [{\citenamefont {Landi}\ and\ \citenamefont
  {Paternostro}(2021)}]{Landi2020}%
  \BibitemOpen
  \bibfield  {author} {\bibinfo {author} {\bibfnamefont {Gabriel~T.}\
  \bibnamefont {Landi}}\ and\ \bibinfo {author} {\bibfnamefont {Mauro}\
  \bibnamefont {Paternostro}},\ }\bibfield  {title} {\enquote {\bibinfo {title}
  {Irreversible entropy production, from quantum to classical},}\ }\href
  {\doibase 10.1103/RevModPhys.93.035008} {\bibfield  {journal} {\bibinfo
  {journal} {Rev. Mod. Phys.}\ }\textbf {\bibinfo {volume} {93}},\ \bibinfo
  {pages} {035008} (\bibinfo {year} {2021})}\BibitemShut {NoStop}%
\bibitem [{\citenamefont {Deffner}\ and\ \citenamefont
  {Bonan{\c{c}}a}(2020)}]{Deffner2020EPL}%
  \BibitemOpen
  \bibfield  {author} {\bibinfo {author} {\bibfnamefont {Sebastian}\
  \bibnamefont {Deffner}}\ and\ \bibinfo {author} {\bibfnamefont {Marcus
  V.~S.}\ \bibnamefont {Bonan{\c{c}}a}},\ }\bibfield  {title} {\enquote
  {\bibinfo {title} {Thermodynamic control {\textemdash}an old paradigm with
  new applications},}\ }\href {\doibase 10.1209/0295-5075/131/20001} {\bibfield
   {journal} {\bibinfo  {journal} {{EPL} (Europhysics Letters)}\ }\textbf
  {\bibinfo {volume} {131}},\ \bibinfo {pages} {20001} (\bibinfo {year}
  {2020})}\BibitemShut {NoStop}%
\bibitem [{\citenamefont {Drude}(1900{\natexlab{a}})}]{Drude1900I}%
  \BibitemOpen
  \bibfield  {author} {\bibinfo {author} {\bibfnamefont {P.}~\bibnamefont
  {Drude}},\ }\bibfield  {title} {\enquote {\bibinfo {title} {{Zur
  Elektronentheorie der Metalle}},}\ }\href {\doibase 10.1002/andp.19003060312}
  {\bibfield  {journal} {\bibinfo  {journal} {Ann. Phys.}\ }\textbf {\bibinfo
  {volume} {306}},\ \bibinfo {pages} {566--613} (\bibinfo {year}
  {1900}{\natexlab{a}})}\BibitemShut {NoStop}%
\bibitem [{\citenamefont {Drude}(1900{\natexlab{b}})}]{Drude1900II}%
  \BibitemOpen
  \bibfield  {author} {\bibinfo {author} {\bibfnamefont {P.}~\bibnamefont
  {Drude}},\ }\bibfield  {title} {\enquote {\bibinfo {title} {{Zur
  Elektronentheorie der Metalle; II. Teil. Galvanomagnetische und
  thermomagnetische Effecte}},}\ }\href {\doibase 10.1002/andp.19003081102}
  {\bibfield  {journal} {\bibinfo  {journal} {Ann. Phys.}\ }\textbf {\bibinfo
  {volume} {308}},\ \bibinfo {pages} {369--402} (\bibinfo {year}
  {1900}{\natexlab{b}})}\BibitemShut {NoStop}%
\bibitem [{\citenamefont {Sommerfeld}(1928)}]{Sommerfeld1928}%
  \BibitemOpen
  \bibfield  {author} {\bibinfo {author} {\bibfnamefont {A.}~\bibnamefont
  {Sommerfeld}},\ }\bibfield  {title} {\enquote {\bibinfo {title} {{Zur
  Elektronentheorie der Metalle auf Grund der Fermischen Statistik}},}\ }\href
  {\doibase 10.1007/BF01391052} {\bibfield  {journal} {\bibinfo  {journal} {Z.
  Phys.}\ }\textbf {\bibinfo {volume} {47}},\ \bibinfo {pages} {1--32}
  (\bibinfo {year} {1928})}\BibitemShut {NoStop}%
\bibitem [{\citenamefont {Bardeen}(1940)}]{bardeen1940}%
  \BibitemOpen
  \bibfield  {author} {\bibinfo {author} {\bibfnamefont {J.}~\bibnamefont
  {Bardeen}},\ }\bibfield  {title} {\enquote {\bibinfo {title} {Electrical
  conductivity of metals},}\ }\href
  {https://aip.scitation.org/doi/10.1063/1.1712751} {\bibfield  {journal}
  {\bibinfo  {journal} {J. Appl. Phys.}\ }\textbf {\bibinfo {volume} {11}},\
  \bibinfo {pages} {88} (\bibinfo {year} {1940})}\BibitemShut {NoStop}%
\bibitem [{\citenamefont {Olmon}\ \emph {et~al.}(2012)\citenamefont {Olmon},
  \citenamefont {Slovick}, \citenamefont {Johnson}, \citenamefont {Shelton},
  \citenamefont {Oh}, \citenamefont {Boreman},\ and\ \citenamefont
  {Raschke}}]{Olmon2012}%
  \BibitemOpen
  \bibfield  {author} {\bibinfo {author} {\bibfnamefont {Robert~L.}\
  \bibnamefont {Olmon}}, \bibinfo {author} {\bibfnamefont {Brian}\ \bibnamefont
  {Slovick}}, \bibinfo {author} {\bibfnamefont {Timothy~W.}\ \bibnamefont
  {Johnson}}, \bibinfo {author} {\bibfnamefont {David}\ \bibnamefont
  {Shelton}}, \bibinfo {author} {\bibfnamefont {Sang-Hyun}\ \bibnamefont {Oh}},
  \bibinfo {author} {\bibfnamefont {Glenn~D.}\ \bibnamefont {Boreman}}, \ and\
  \bibinfo {author} {\bibfnamefont {Markus~B.}\ \bibnamefont {Raschke}},\
  }\bibfield  {title} {\enquote {\bibinfo {title} {Optical dielectric function
  of gold},}\ }\href {\doibase 10.1103/PhysRevB.86.235147} {\bibfield
  {journal} {\bibinfo  {journal} {Phys. Rev. B}\ }\textbf {\bibinfo {volume}
  {86}},\ \bibinfo {pages} {235147} (\bibinfo {year} {2012})}\BibitemShut
  {NoStop}%
\bibitem [{\citenamefont {Yang}\ \emph {et~al.}(2015)\citenamefont {Yang},
  \citenamefont {D'Archangel}, \citenamefont {Sundheimer}, \citenamefont
  {Tucker}, \citenamefont {Boreman},\ and\ \citenamefont {Raschke}}]{Yang2015}%
  \BibitemOpen
  \bibfield  {author} {\bibinfo {author} {\bibfnamefont {Honghua~U.}\
  \bibnamefont {Yang}}, \bibinfo {author} {\bibfnamefont {Jeffrey}\
  \bibnamefont {D'Archangel}}, \bibinfo {author} {\bibfnamefont {Michael~L.}\
  \bibnamefont {Sundheimer}}, \bibinfo {author} {\bibfnamefont {Eric}\
  \bibnamefont {Tucker}}, \bibinfo {author} {\bibfnamefont {Glenn~D.}\
  \bibnamefont {Boreman}}, \ and\ \bibinfo {author} {\bibfnamefont {Markus~B.}\
  \bibnamefont {Raschke}},\ }\bibfield  {title} {\enquote {\bibinfo {title}
  {Optical dielectric function of silver},}\ }\href {\doibase
  10.1103/PhysRevB.91.235137} {\bibfield  {journal} {\bibinfo  {journal} {Phys.
  Rev. B}\ }\textbf {\bibinfo {volume} {91}},\ \bibinfo {pages} {235137}
  (\bibinfo {year} {2015})}\BibitemShut {NoStop}%
\bibitem [{\citenamefont {Reichl}(2016)}]{reichl}%
  \BibitemOpen
  \bibfield  {author} {\bibinfo {author} {\bibfnamefont {L.~E.}\ \bibnamefont
  {Reichl}},\ }\href@noop {} {\emph {\bibinfo {title} {A modern course in
  statistical physics}}}\ (\bibinfo  {publisher} {Wiley-VCH},\ \bibinfo
  {address} {New York},\ \bibinfo {year} {2016})\BibitemShut {NoStop}%
\bibitem [{\citenamefont {Kubo}\ \emph {et~al.}(2012)\citenamefont {Kubo},
  \citenamefont {Toda},\ and\ \citenamefont {Hashitsume}}]{kubo2012}%
  \BibitemOpen
  \bibfield  {author} {\bibinfo {author} {\bibfnamefont {R.}~\bibnamefont
  {Kubo}}, \bibinfo {author} {\bibfnamefont {M.}~\bibnamefont {Toda}}, \ and\
  \bibinfo {author} {\bibfnamefont {N.}~\bibnamefont {Hashitsume}},\ }\href
  {https://books.google.com.br/books?id=cF3wCAAAQBAJ&dq=kubo+statistical+2&source=gbs_navlinks_s}
  {\emph {\bibinfo {title} {Statistical physics II: nonequilibrium statistical
  mechanics}}},\ Vol.~\bibinfo {volume} {31}\ (\bibinfo  {publisher} {Springer
  Science \& Business Media},\ \bibinfo {year} {2012})\BibitemShut {NoStop}%
\bibitem [{\citenamefont {Ashcroft}\ and\ \citenamefont
  {Mermin}(1976)}]{Ashcroft1976}%
  \BibitemOpen
  \bibfield  {author} {\bibinfo {author} {\bibfnamefont {Neil~W}\ \bibnamefont
  {Ashcroft}}\ and\ \bibinfo {author} {\bibfnamefont {N~David}\ \bibnamefont
  {Mermin}},\ }\href@noop {} {\emph {\bibinfo {title} {Solid state physics}}}\
  (\bibinfo  {publisher} {Saunders College Publishing},\ \bibinfo {year}
  {1976})\BibitemShut {NoStop}%
\bibitem [{Note1()}]{Note1}%
  \BibitemOpen
  \bibinfo {note} {Note, however, that it has been shown that the entropy
  production, $\protect \overline {\Sigma }=\DOTSI \intop \ilimits@
  _{0}^{t}dt'\protect \tmspace +\thinmuskip {.1667em}\protect \mathaccentV
  {dot}05F{\protect \overline {\Sigma }} $, itself remains positive at all
  times \cite {Bonanca2020a,Bonanca2021}.}\BibitemShut {Stop}%
\bibitem [{\citenamefont {Zon}\ and\ \citenamefont {Cohen}(2003)}]{zon2003}%
  \BibitemOpen
  \bibfield  {author} {\bibinfo {author} {\bibfnamefont {R.~van}\ \bibnamefont
  {Zon}}\ and\ \bibinfo {author} {\bibfnamefont {E.~G.~D.}\ \bibnamefont
  {Cohen}},\ }\bibfield  {title} {\enquote {\bibinfo {title} {Stationary and
  transient work-fluctuation theorems for a dragged brownian particle},}\
  }\href {\doibase 10.1103/PhysRevE.67.046102} {\bibfield  {journal} {\bibinfo
  {journal} {Phys. Rev. E}\ }\textbf {\bibinfo {volume} {67}},\ \bibinfo
  {pages} {046102} (\bibinfo {year} {2003})}\BibitemShut {NoStop}%
\bibitem [{\citenamefont {Jim\'enez-Aquino}(2010)}]{aquino2010}%
  \BibitemOpen
  \bibfield  {author} {\bibinfo {author} {\bibfnamefont {J.~I.}\ \bibnamefont
  {Jim\'enez-Aquino}},\ }\bibfield  {title} {\enquote {\bibinfo {title}
  {Entropy production theorem for a charged particle in an eletromagnetic
  field},}\ }\href {\doibase 10.1103/PhyRevE.82.051118} {\bibfield  {journal}
  {\bibinfo  {journal} {Phys. Rev. E}\ }\textbf {\bibinfo {volume} {82}},\
  \bibinfo {pages} {051118} (\bibinfo {year} {2010})}\BibitemShut {NoStop}%
\bibitem [{\citenamefont {Chernyak}\ \emph {et~al.}(2006)\citenamefont
  {Chernyak}, \citenamefont {Chertkov},\ and\ \citenamefont
  {Jarzysnki}}]{Chernyak2006}%
  \BibitemOpen
  \bibfield  {author} {\bibinfo {author} {\bibfnamefont {V.~Y.}\ \bibnamefont
  {Chernyak}}, \bibinfo {author} {\bibfnamefont {M.}~\bibnamefont {Chertkov}},
  \ and\ \bibinfo {author} {\bibfnamefont {C.}~\bibnamefont {Jarzysnki}},\
  }\bibfield  {title} {\enquote {\bibinfo {title} {Path-integral analysis of
  fluctuation theorems for general langevin processes},}\ }\href {\doibase
  10.1088/1742-5468/2006/08/P08001} {\bibfield  {journal} {\bibinfo  {journal}
  {J. Stat, Mech.}\ }\textbf {\bibinfo {volume} {2006}},\ \bibinfo {pages}
  {P08001} (\bibinfo {year} {2006})}\BibitemShut {NoStop}%
\bibitem [{\citenamefont {Imparato}\ and\ \citenamefont
  {Peliti}(2006)}]{Imparato2006}%
  \BibitemOpen
  \bibfield  {author} {\bibinfo {author} {\bibfnamefont {A.}~\bibnamefont
  {Imparato}}\ and\ \bibinfo {author} {\bibfnamefont {L.}~\bibnamefont
  {Peliti}},\ }\bibfield  {title} {\enquote {\bibinfo {title} {Fluctuation
  relations for a driven brownian particle},}\ }\href {\doibase
  10.1103/PhysRevE.74.026106} {\bibfield  {journal} {\bibinfo  {journal} {Phys.
  Rev. E}\ }\textbf {\bibinfo {volume} {74}},\ \bibinfo {pages} {026106}
  (\bibinfo {year} {2006})}\BibitemShut {NoStop}%
\bibitem [{\citenamefont {Deffner}\ \emph {et~al.}(2011)\citenamefont
  {Deffner}, \citenamefont {Brunner},\ and\ \citenamefont
  {Lutz}}]{Deffner2011EPL}%
  \BibitemOpen
  \bibfield  {author} {\bibinfo {author} {\bibfnamefont {S.}~\bibnamefont
  {Deffner}}, \bibinfo {author} {\bibfnamefont {M.}~\bibnamefont {Brunner}}, \
  and\ \bibinfo {author} {\bibfnamefont {E.}~\bibnamefont {Lutz}},\ }\bibfield
  {title} {\enquote {\bibinfo {title} {Quantum fluctuation theorems in the
  strong damping limit},}\ }\href {\doibase 10.1209/0295-5075/94/30001}
  {\bibfield  {journal} {\bibinfo  {journal} {{EPL} (Europhysics Letters)}\
  }\textbf {\bibinfo {volume} {94}},\ \bibinfo {pages} {30001} (\bibinfo {year}
  {2011})}\BibitemShut {NoStop}%
\bibitem [{\citenamefont {Pal}\ and\ \citenamefont
  {Deffner}(2020)}]{Pal2020NJP}%
  \BibitemOpen
  \bibfield  {author} {\bibinfo {author} {\bibfnamefont {P~S}\ \bibnamefont
  {Pal}}\ and\ \bibinfo {author} {\bibfnamefont {Sebastian}\ \bibnamefont
  {Deffner}},\ }\bibfield  {title} {\enquote {\bibinfo {title} {Stochastic
  thermodynamics of relativistic brownian motion},}\ }\href {\doibase
  10.1088/1367-2630/ab9ce6} {\bibfield  {journal} {\bibinfo  {journal} {New J.
  Phys.}\ }\textbf {\bibinfo {volume} {22}},\ \bibinfo {pages} {073054}
  (\bibinfo {year} {2020})}\BibitemShut {NoStop}%
\bibitem [{\citenamefont {Risken}(1996)}]{Risken}%
  \BibitemOpen
  \bibfield  {author} {\bibinfo {author} {\bibfnamefont {H.}~\bibnamefont
  {Risken}},\ }\href@noop {} {\emph {\bibinfo {title} {The Fokker-Planck
  equation}}}\ (\bibinfo  {publisher} {Springer-Verlag},\ \bibinfo {address}
  {Berlin},\ \bibinfo {year} {1996})\BibitemShut {NoStop}%
\bibitem [{\citenamefont {Ferrari}(2003)}]{ferrari2003}%
  \BibitemOpen
  \bibfield  {author} {\bibinfo {author} {\bibfnamefont {L.}~\bibnamefont
  {Ferrari}},\ }\bibfield  {title} {\enquote {\bibinfo {title} {Heavy (or
  large) ions in a fluid in an electric field: the fundamental solution of the
  fokker-planck equation and related questions},}\ }\href {\doibase
  10.1063/1.1574779} {\bibfield  {journal} {\bibinfo  {journal} {J. Chem.
  Phys}\ }\textbf {\bibinfo {volume} {118}},\ \bibinfo {pages} {11092}
  (\bibinfo {year} {2003})}\BibitemShut {NoStop}%
\end{thebibliography}
%

\end{document}